\documentclass{rnaastex}

\begin{document}

\title{Observation of ``Topological'' Microflares in the Solar Atmosphere}

\correspondingauthor{Yurii V. Dumin}
\email{dumin@pks.mpg.de, dumin@yahoo.com}

\author[0000-0002-9604-5304]{Yurii V. Dumin}

\altaffiliation{Currently in the Max Planck Institute for
                the Physics of Complex Systems, Dresden, Germany}

\affiliation{Sternberg Astronomical Institute of
             Lomonosov Moscow State University, \\
             13 Universitetsky prosp., Moscow 119234, Russia}

\affiliation{Space Research Institute of Russian Academy of Sciences}

\author{Boris V. Somov}

\affiliation{Sternberg Astronomical Institute of
             Lomonosov Moscow State University, \\
             13 Universitetsky prosp., Moscow 119234, Russia}

\keywords{magnetic reconnection --- Sun: flares}

\section{Observational Data}

\begin{figure}[h!]
\begin{center}
\includegraphics[scale=0.95,angle=0]{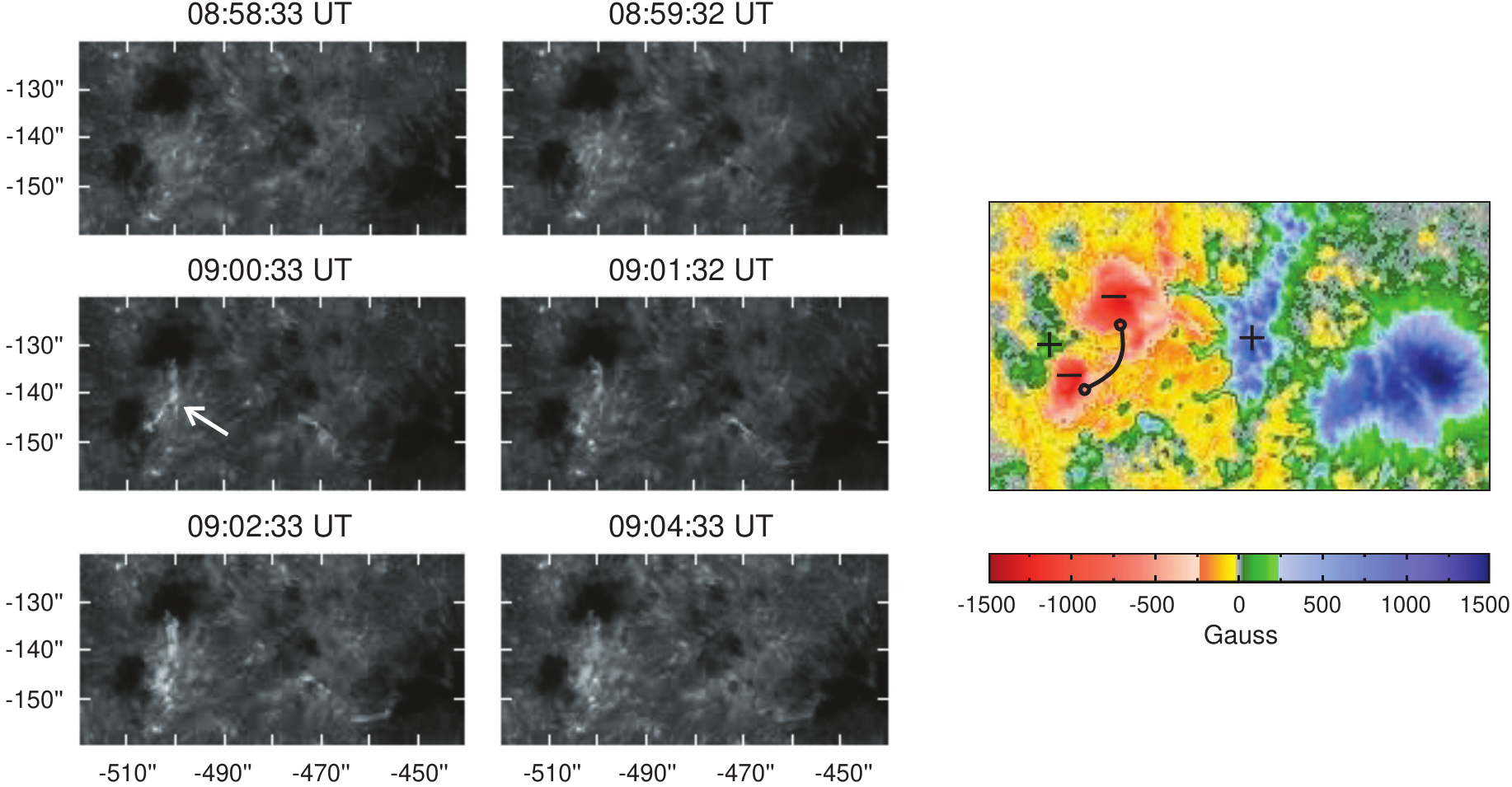}
\caption{Left: Frames of emission in the \ion{Ca}{2} line taken by
\textit{Hinode} SOT on 1~October 2014 at a few successive instants of time.
Right: Magnetogram of the same region recorded by \textit{SDO} HMI at
09:00:00~TAI.
(Courtesy of NASA/SDO and the AIA, EVE, and HMI science teams.)
\label{fig:1}}
\end{center}
\end{figure}

We report here on observation of the unusual kind of solar microflares,
presumably associated with the so-called ``topological trigger''
of magnetic reconnection, which was theoretically suggested long time ago
by \citet{Gorbachev_88} but has not been clearly identified so far by
observations.
The corresponding example is shown in the left-hand side of
Figure~\ref{fig:1}: these frames were recorded by the Solar Optical
Telescope (SOT) onboard \textit{Hinode} satellite \citep{Kosugi_07,Tsuneta_08}
in the chromospheric line~\ion{Ca}{2}.%
\footnote{%
The data were taken from the Hinode QL Movie Archive at
\url{http://hinode.nao.ac.jp/en/for-researchers/qlmovies/}}
At first sight, the bright loop flaring at 09:00:33~UT is just a magnetic
field line connecting two sunspots.
However, a closer inspection of the \textit{SDO} HMI magnetogram
\citep{Pesnell_12} for the respective instant%
\footnote{%
The data were taken from the Joint Science Operations Center (JSOC) at
\url{http://jsoc.stanford.edu/HMI/hmiimage.html}}
shows that the arc is anchored in the regions of the same polarity near
the sunspot boundaries.
So, it cannot be an ordinary magnetic-field line connecting the opposite
magnetic poles.

Yet another peculiar feature is that this arc flashes almost instantly as
a thin strip and then begins to expand and decay, while typical
the chromospheric flares in \ion{Ca}{2} line are much wider and propagate
progressively in space.

\section{Theoretical Interpretation}

A qualitative explanation of the observed phenomenon can be given by the
above-mentioned model of topological trigger.
Namely, there are such configurations of the magnetic sources (e.g. point-like
magnetic poles) on the surface of photosphere that their tiny displacements
result in the formation and fast motion of a 3D null point along the arc
located well above the plane of the sources.
So, such a null point can quickly ignite a magnetic reconnection along
the entire its trajectory.
Pictorially, this can be presented as flipping the so-called two-dome
magnetic-field structure (which is just the reason why such mechanism
of reconnection was called topological).
The general mathematical criteria for its onset can be found in
paper by~\citet{Gorbachev_88}, while our recent numerical simulations will
be published elsewhere.
In brief, the most important prerequisite for the development of topological
instability in the two-dome structure is a cruciform arrangement of
the magnetic sources in its base.
As can be seen in the right panel of Figure~\ref{fig:1}, this condition is
really satisfied in the case under consideration.

Let us mention that previous attempts to identify the topological trigger
in the solar atmosphere were concentrated mostly on the large flares
\citep[e.g.][]{OreshinaA_12,OreshinaI_09}.
As a result, it was found that the magnetic-field configuration required
for the development of topological trigger is sometimes formed just before
the onset of the powerful flares.
Unfortunately, it was impossible to identify the trajectory of the 3D
null point, because the entire picture was strongly obscured by the intense
electric currents generated immediately in the spot of reconnection.
The advantage of the present study is that
(1)~we analyzed the microflares, which are governed mostly by the external
(photospheric) magnetic sources, and
(2)~we employed observations in the chromospheric \ion{Ca}{2}~line,
which is emitted by a moderately heated plasma and, thereby, can trace
the initial stage of reconnection.

\acknowledgments

We are grateful to A.~V. Getling, I.~V. Oreshina, and A.~V. Oreshina for
valuable discussions and advices.
\textit{Hinode} is a Japanese mission developed and launched by ISAS/JAXA,
with NAOJ as domestic partner and NASA and STFC (UK) as international
partners.
It is operated by these agencies in co-operation with ESA and NSC (Norway).

\end{document}